\documentstyle[aps,psfig]{revtex}

\begin{document}
\draft
\title{Constraint Molecular Dynamics 
approach to Fermionic systems.}

\author{Massimo Papa$^{a,b)}$\footnote{e-mail: papa@lns.infn.it}, 
Toshiki Maruyama$^{a,c)}$\footnote{e-mail: maru@hadron02.tokai.jaeri.go.jp} 
and Aldo Bonasera$^{a)}$\footnote{e-mail: bonasera@lns.infn.it} }

\address{
{\it a)Istituto Nazionale Fisica Nucleare-Laboratorio Nazionale 
del Sud, Via Santa Sofia 44,
95123 Catania, Italy}\\
{\it b) Istituto Nazionale Fisica Nucleare-Sezione di Catania, 
Corso Italia 57, 95129 Italy}\\
{\it c)Advanced Science Research Center, Japan Atomic Energy Research Institute,
Tokai, Ibaraki, 319-1195, Japan}}

\maketitle

\begin{abstract}
We propose a Constraint Molecular Dynamics model for Fermionic system.
In this approach 
the equations of motion of wave packets for the nuclear 
many-body problem are solved by imposing that the one-body occupation probability 
$\overline{f}(r,p,t)$  can assume only values less or equal to 1. 
This condition reflects the Fermionic nature of the studied systems and 
it is implemented with a fast algorithm which allows also the study 
of the heaviest colliding system.
The parameters of the model have been chosen to reproduce the average 
binding energy and radii of nuclei in the mass region $A=30\sim 208$.
Some comparison to data is given.

\end{abstract}

\section{Introduction}

Heavy ion collisions in the medium energy region have been described
in a large variety of semiclassical approaches to the many-body problem.
As it is well known the one-body  semiclassical transport models 
like the Boltzman-Northeim-Vlasov (BNV)\cite{bon} and 
Vlasov-Uehling-Uhlenbeck (VUU)\cite{bert} are not suited to describe 
processes in which a large number of final fragments are produced. 
 This is due to the fact that the correlations treated in the one-body approach 
 are not able to describe the large fluctuations which develops 
 in a multifragmentation process.

This difficulty can be solved by adopting more suitable 
treatments of the $N$-body problem like molecular dynamics.
Several molecular dynamics models have been developed up to now \cite{bonasera2000}. 
In the quantum molecular dynamics (QMD)\cite{aich} the $N$-body wave function
is expressed through a direct product of wave packets each of which satisfies 
the minimum uncertainty relation $\sigma_{r}\sigma_{p}=\hbar/2$, 
where $\sigma_{r}$ and $\sigma_{p}$ represent the dispersion of the corresponding 
Wigner transform in configuration and momentum space respectively.  

The Fermionic nature of the nuclear many-body problem has been considered 
in the Fermionic molecular dynamics \cite{feld} and more recently
in the antisymmetrized molecular dynamics (AMD) \cite{ono}. 
In these models, 
the wave function of the system is expressed as 
a single Slater determinant of $N$ wave packets.
In this way the Fermionic nature of the system is preserved.
In particular in the AMD approach two-body collisions are introduced 
in the ``physical coordinates'' which are obtained by a canonical transformation from 
the parameter coordinates of wave-packets \cite{ono}.
The nucleon-nucleon collision is only allowed when an inverse transformation from 
a newly chosen ``physical coordinate'' (the final state of the collision) 
into the corresponding parameter coordinate exists. 
This condition is understood as a stochastic change of the state from 
a single Slater determinant into another, both of which have occupation 
probability 1 or 0. 
Due to the four dimensional matrix element of two-body interaction, 
the CPU time  necessary to work out calculations for systems with 
total mass larger than 200 is very large for practical studies.

On the other hand the QMD calculations are much easier to carry out because
they need in general only double-fold loops to calculate two-body interactions.
But it is obvious that the Fermionic feature is lacking in the QMD model.
 Two-body collisions are also introduced phenomenologically.
The Pauli blocking of the final state is usually checked in a 
similar manner to the collision term in the BNV model. 
The two-body collisions with Pauli blocking have 
some effects to maintain the Fermionic feature of the system.
However, 
in the ground states or in low energy reaction phenomena, 
two-body collisions are absent or very rare.
Even if the initial state is in good agreement with the phase space distribution
of a Fermionic system, the time evolution by the classical equations of motion surely 
breaks the initial distribution which evolves to a classical Boltzmann one.

To compensate this shortcoming, the Pauli potential is introduced by several 
authors to mimic the Fermionic features 
\cite{wil,Dorso,Mar97}.
This phenomenological potential forbids nucleons of the same spin and 
isospin from coming close to each other in the phase space.
Although the Pauli potential gives us some good results such as 
stable ground states, with energies in good agreement with experiments and
saturation properties of nuclear matter,
it also gave some undesirable by-products, for instance,
spurious repulsion in the collision problem.

In the present work we propose a new molecular dynamics model: 
the Constraint Molecular Dynamics (CoMD), which aims 
to overcome the above mentioned limitations. 
In particular we want: \\
i) to describe the Fermionic nature of the $N$-body system with
the more general condition that the occupation probability 
$\overline{f}\leq 1$;\\
ii) to realize a model for which the computational time 
is short enough to allow the study of the heaviest systems.

\section{The Model}
\subsection{Theoretical framework}

In QMD, each nucleon state is represented by a Gaussian
wave-function of width $\sigma_r$,
\begin{equation}
\phi_{i}({\bf r}) = \frac{1}{(2\pi\sigma_r^{2})^{3/4}} \exp \left[
                - \frac{({\bf r} - \langle{\bf r}_i\rangle)^2}{4{\sigma_r}^2} +
             \frac{i}{\hbar} {\bf r} \cdot \langle{\bf p}_i\rangle
\right],
\end{equation}
where $\langle{\bf r}_i\rangle$ and $\langle{\bf p}_i\rangle$ are the 
centers of position
and momentum of $i$-th nucleon, respectively.
The total wave-function is assumed to be a direct product of
these wave-functions. Therefore the N-body distribution function
is the direct product of the single particle distribution functions
$f_{i}$. The $f_{i}$ are obtained by the Wigner 
transform of the wave-functions $\phi_{i}$ and the one-body distribution
function is given by:
\begin{eqnarray}
  f({\bf r},{\bf p}) &=&  \sum_i { f_i({\bf r},{\bf p}) },\\
f_i({\bf r},{\bf p}) &=&  {1 \over \pi^{3}\hbar^{3}} \cdot
              \exp\left[-{({\bf r}-\langle{\bf r}_i\rangle)^2\over 
2{\sigma_r}^2}
            -{2{\sigma_r}^2({\bf p}-\langle{\bf p}_i\rangle)^2\over
\hbar^2}\right].
\end{eqnarray}

We note that since $\sigma_{r}$ is a real number in the QMD approach
the distribution function $f_{i}$ produce the minimum uncertainty
relation $\sigma_{r}\sigma_{p}=\hbar/2$ in one-body phase space.

In this paper we extend this kind of representation. We in fact 
take the dispersion in momentum $\sigma_p$ as
a parameter as well as that in coordinate space allowing for the more
general
uncertainty condition
$\sigma_p\sigma_r\ge\hbar/2$. Therefore we write the one-body distribution 
function as:
\begin{eqnarray}
f_i({\bf r},{\bf p}) &=&  \frac{1}{(2\pi\sigma_r\sigma_p)^3} \cdot
      \exp\left[-{({\bf r}-\langle{\bf r}_i\rangle)^2\over 
2{\sigma_r}^2}
                -{({\bf p}-\langle{\bf p}_i\rangle)^2\over 
2{\sigma_p}^2}\right].
\end{eqnarray}

This extension implies that each single particle wave function $\phi_{i}$
can be now represented by an incoherent 
superimposition of a large number of minimum uncertainty wave-packets.
 
The equations of motion of $\langle{\bf r}_i\rangle$ and $\langle{\bf 
p}_i\rangle$
are derived using the time-dependent variational principle which gives:
\begin{equation}
\dot{\langle{\bf r}_i\rangle} =   \frac{\partial H}{\partial
\langle{\bf p}_i\rangle},
\;\;\;\;
\dot{\langle{\bf p}_i\rangle} = - \frac{\partial H}{\partial
\langle{\bf r}_i\rangle}.
\label{eq4}
\end{equation}
The Hamiltonian $H$ consists of the kinetic energy
and the two-body effective interactions:
\begin{eqnarray}
H&=&\sum_i {\langle{\bf p}_i\rangle^2\over 2m} +{1\over 2}\sum_{i,j\neq i}
\overline{V}_{ij}\\
\overline{V}_{ij} &\equiv&
   \int d^3r_i\;d^3r_j\; \rho_i({\bf r}_i)\rho_j({\bf r}_j)V_{ij},\label{eq3}\\
\rho_i &\equiv& \int d^3p\;f_i({\bf r,p}),
\end{eqnarray}
where $V_{ij}$ represents two body interaction between particles $i$ and 
$j$
and $\rho_i$ represents the density distribution of nucleon $i.$
One should note that the kinetic energy coming from the momentum dispersion
$3{\sigma_p}^2/2m$ is omitted as a spurious constant term.

\subsection{Effective interaction}

The nucleon-nucleon interaction has been described through the sum
of several terms:
\begin{eqnarray}
V_{ij}&=&V^{\rm vol}+V^{(3)}+V^{\rm asy}+V^{\rm sup}+V^{\rm Coul}\\
V^{\rm vol}&=&{t_{0} \over \rho_{0}}
\delta({\bf r}_{i}-{\bf r}_{j})\\
V^{(3)}&=&t_{3}
\left[{1 \over \rho_{0}}
\delta({\bf r}_{i}-{\bf r}_{j})\right]
^{7/6}\\
V^{\rm asy}&=&{a_{\rm sym} \over \rho_{0}}\delta({\bf r}_{i}-
{\bf r}_{j}) [2\delta_{\tau_i, \tau_j}-1] \\
V^{\rm sup}&=&C_{s}\nabla^{2}_{\langle{\bf r}_{i}\rangle}
\delta({\bf r}_{i}-{\bf r}_{j})\\
V^{\rm Coul}&=&{e^{2} \over |{\bf r}_{i}-{\bf r}_{j}|}
\end{eqnarray}
In the above relations the power operator $[..]^{7/6}$ and $\nabla^{2}$ act
after the folding operations indicated in Eq.~(\ref{eq3}). 
The $\tau_{i}$ indicates the isospin degree of freedom. 
The $V^{\rm vol}$ and the $V^{(3)}$ terms represent the two-body and 
the so called three-body term of the well known Skyrme interaction. 
The value of $t_{0}$ and $t_{3}$ have been fixed to $-356$ MeV and $303$ MeV. 
These values reproduce the saturation density $\rho_{0}$ and binding energy 
for symmetric nuclear matter with a compressibility of $200$ MeV.
The third term represent the asymmetry term with $a_{\rm sym}=32$ MeV.
The parameter $C_{s}$ of the ``surface'' term $V^{\rm surf}$ is
chosen through a global fit procedure to reproduce, on the average,
the radii
and the binding energies for nuclei with mass-ranging from 30 to 208.

\subsection{Numerical methods and the Constraint}

The set of equations expressed in Eq.~(\ref{eq4}) have been solved
using a fourth-order Runge-Kutta method coupled with two numerical algorithms
which aim to take into account the effects of the residual interaction
and the Fermionic nature of the many-body problem we are studying.

One consists of the
usual two-body elastic collisions which mimics the effect
of the short range repulsive residual interaction together with the
stochastic change of the phase-space distribution with
Pauli blocking in the final states.
The isospin dependent parameterization of the nucleon-nucleon
elastic angular distribution
together with the concept of mean-free
path have been used to compute the collision probability per unit of
time \cite{bon}.
The Pauli blocking factor, which is related to the constraint, is discussed later.

The other algorithm constraints at each time step
the following quantities:
\begin{eqnarray}
\overline{f}_{i} &\leq& 1\ \ \ \ \ \  \hbox{(for all $i$)}  \label{eq8} 
\\
\overline{f}_{i} &\equiv&
 \sum_j \delta_{\tau_{i},\tau_{j}}
  \delta_{s_{i},s_{j}}\int_{h^{3}}f_j({\bf r}, {\bf p};
  \langle {\bf r}_j\rangle,\langle {\bf p}_j\rangle)\;d^3r\;d^3p
\end{eqnarray}
The coordinate $s_{i}$ represent the nucleon spin projection quantum 
number.
The integral is performed on an
hypercube
of volume $h^{3}$ in the phase-space
centered around the point $(\langle {\bf r}_i\rangle,\langle {\bf 
p}_i\rangle)$
with size
$\sqrt{{2\pi\hbar \over \sigma_{r}\sigma_{p}}}\sigma_{r}$ and
$\sqrt{{2\pi\hbar \over \sigma_{r}\sigma_{p}}}\sigma_{p}$ in the  
$r$ and $p$ spaces respectively.
The quantities $\overline{f}_i$ can therefore be interpreted like an
occupation probability of the one-body phase-space around the point
of coordinate $(\langle {\bf r}_i\rangle, \langle {\bf p}_i\rangle)$.
In general more than $60\%$ of the occupation probability 
$\overline{f}_i$
arises from the contribution $f_i$ of the particle $i$ itself.
To realize the constraint expressed trough the relation (\ref{eq8})
the following procedure has been used:

\noindent
At each time step and for each particle $i$ an ensemble $K_i$ of nearest 
identical particles (including the particle $i$) is determined within the 
distances $3\sigma_r$ and $3\sigma_p$ in the phase space.
If the phase space occupation $\overline{f}_i$  has a value greater than 1
we perform many body elastic scattering among the ensemble $K_i$
in order to reduce the phase space occupation.
To reduce the CPU time the many body scattering process
is practically done by
a series of two-body scatterings.

The Pauli blocking probability 
for the usual two-body collision process 
is given as:
\begin{eqnarray}
P_{\rm block} &=& 1-(1-S_{i})(1-S_{j}), \\
S_{i} &\equiv&
  {\rm min}\left(1, {\overline{f}_i-{\overline{f}_i}^i \over
1-{\overline{f}_i}^i}
	  \right),
\label{eq9} \\
{\overline{f}_i}^i &\equiv&
 \int_{h^{3}}
 f_{i}({\bf r},{\bf p};\langle {\bf r}_{i}\rangle,\langle {\bf p}_{i}\rangle)
 \;d^3r\;d^3p.
\end{eqnarray}
The quantity $S_i$ in Eq.~(\ref{eq9}) is the ``effective occupation''
of phase space at the position of particle $i$. 
In fact in the definition of $S_i$ we are subtracting the contribution of
the particle $i$ itself. 

We stress that the constraint acts in a way complementary to the 
collision term.
In fact particles of low momenta are strongly effected by the constraint in
such a way to avoid that the distribution becomes a classical one.
On the contrary the collision term is especially important for particles
located at high relative momenta.

\section{Calculations and comparison with other models}

As an example we compare our results on the isotope distribution in 
$^{40}{\rm Ca}+^{40}{\rm Ca}$ at 35 MeV/nucleon with the experiment \cite{nat}
and to the result of QMD.
Finally we compare CoMD results to experimental data on 
central Au+Au collisions \cite{ref17}.

\subsection{Initialization}

The ground state configuration of the nuclei is obtained using a 
modified cooling procedure.
The nucleon are first distributed in a sphere of radius $1.4\times 
A^{1/3}$ fm
in configuration space and in a sphere of radius $P_{F}^{\rm nm}$
(Fermi momentum for infinite nuclear matter) in momentum space.
The equations of motion with friction terms
are solved coupled with the constraint. 
At each time step if the value of $\overline{f}_i$
is greater than 1 the momenta of the particles
belonging to the ensemble $K_i$ are scaled by a factor 1.02.
If $\overline{f}_i$ is less than 1 the scale factor is set to 0.98.
With this procedure the total energy 
and the radius $R$ of the nuclei will reach some stationary values.
If the deviations of these values from the experimental ones are within 
$7\%$,
a check is done on the stability.
At this stage the friction term for the ``cooling'' is switched off and
the time dependence of the nuclear radius $R$ is checked.
If $R$ is stable at least for 1000 fm/$c$ (inside the confidence level), 
this initial condition is 
accepted.
In Fig.~1 we show $R$ as function of time for two typical
``good events'' representing the $^{40}{\rm Ca}$ and $^{208}{\rm 
Pb}$ nuclei.
The binding energies are $-8.2$ and $-8.4$ MeV respectively.
The reason of the stability in the CoMD case is the constraint.  

\begin{figure}
\psfig{file=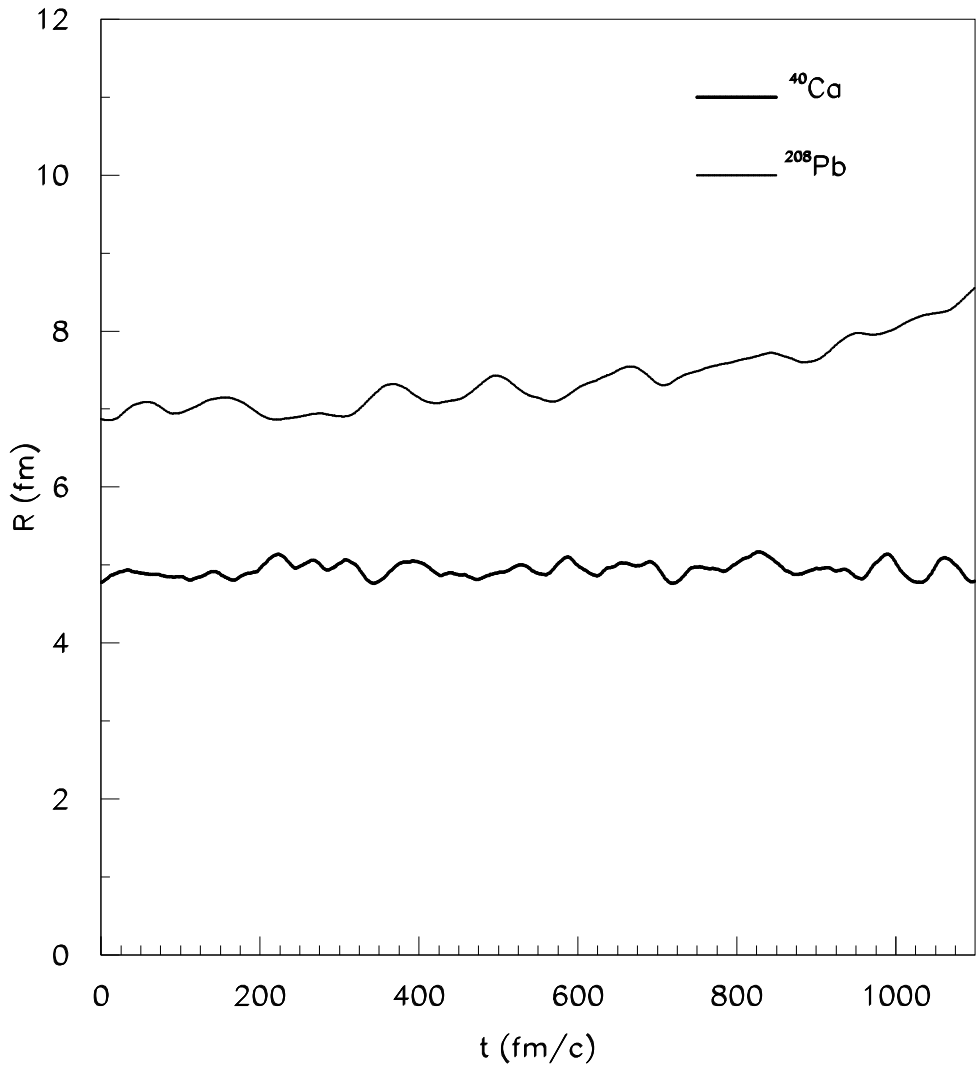,width=0.8\textwidth}
\caption{
 Nuclear radii $R$ as function of time 
for typical ground state configurations of
$^{40}Ca$ and $^{208}{\rm Pb}$ nuclei.
}
\end{figure}

In fact without constraint, the
initial Fermionic distribution will evolve to a classical one.
This means that during the interaction several particles will
tend to have low relative momenta
trying to reach a stable classical configuration
(a classical ground state would be a 
solid) while some other particles (as due to the energy conservation)
will acquire momenta large enough to overcome
the barrier and leave the system.  This does not happen in CoMD since 
the constraint maintains the Fermionic 
nature of the system.

The results shown in Fig.~1 are obtained by setting $\sigma_{r}=1.3$ fm,
$\sigma_{p}/\hbar=0.47$ fm$^{-1}$ and  $C_{s}=-2$ MeV$\cdot$fm$^{2}$.  
The value of the surface term might be surprising at first, 
but we must notice that part of the surface
term comes automatically when using Gaussian.
We observe also that using this procedure with the above set of
parameters the effective average kinetic energy per nucleon $K$
(the kinetic energy regarding the centroid of the Gaussian wave packet)
is very close to the value obtained in a Thomas Fermi model (about 20
MeV)
This feature is instead lost in most of the molecular dynamics models
while it is a peculiarity of the proposed one.

In fact in the present approach
 for a fixed average density and 
effective interaction, the ratio
between  kinetic energy  and the potential one
crucially depends on $\sigma_{p}$ which can
be varied to values greater than $\hbar$/$2\sigma_{r}$ (see Sec.II). 
In particular $\sigma_{p}$ determines the filling of the
phase-space , and therefore, 
the constrained value of $\overline{f}$ will be
automatically reached
with a value of $K$ strongly depending on $\sigma_{p}$.

This feature will allow us, in the global fit
procedure, to find a set of parameters which reproduce
the nuclear binding energies, the radii and
the ratio between $K$ and the potential energy
close to the value of the Thomas Fermi model. 

\subsection{The Nucleus-Nucleus Collision}

In this section we show
some results concerning
the $^{40}{\rm Ca}+^{40}{\rm Ca}$ collision at $E_{\rm lab}=35$
MeV/nucleon.
In Fig.~2 we display the time dependence of $\overline{f}_i$
for all particle index $i$ in a typical event with zero impact
parameter.
The upper and the bottom parts show the results of QMD
and CoMD models, respectively, with exactly the same initial condition.
The difference between the two cases is quite clear.
In the QMD case after a short time (about 50 fm/$c$)
the values of the $\overline{f}_i$ for most of the particles start to
be greater than 1 and are larger than 2 after some hundreds of fm/$c$.
This result is easily understood.
In the QMD case, which has no constraint regarding the Fermionic nature
except for the two-nucleon collision process,
the momentum distribution tends to the classical limit.
Obviously in the CoMD case, due to the constraint,
the phase space occupation $\overline{f}_i$ is
on the average less than 1 at each time step.
We note a good uniformity of $\overline{f}_i$ as function
of the particle index  $i$ and time $t$.
These results obviously affect the collision rate $r_{c}$.

\begin{figure}
\psfig{file=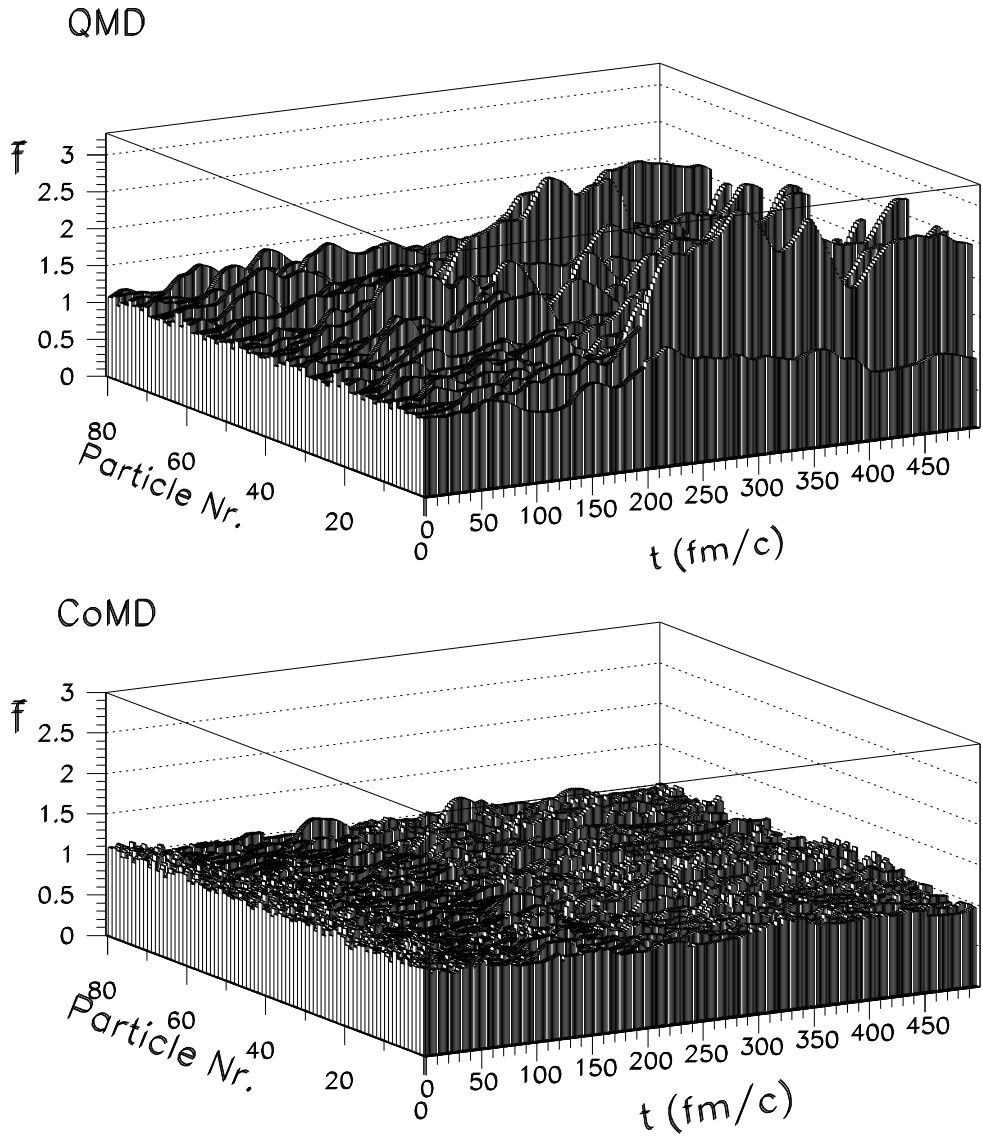,width=0.8\textwidth}
\caption{
 Two-dimensional histograms showing the occupation
probability $\overline{f_{i}}$ as  function of the particle
index $i$ and time for the system under study.
The upper panel refers to the case QMD case while
the lower one is relative to the CoMD  model. 
}
\end{figure}

In Fig.~3 we show
$r_{c}$ as function of $t$ for central collision for both the QMD and the CoMD.
In the QMD case the collision rate steeply increases and decreases
around 40 fm/$c$.
This behavior can be explained as follows:
although we use the same initial condition as CoMD, there appear
some fluctuation of occupation number $\overline{f}_i$ due to the
classical nature of QMD.
Therefore two-nucleon collisions are easily to occur at the beginning
compared to CoMD.
After about 30 fm/$c$, however, when the two nuclei start to overlap
the $\overline{f}_i$ increases spuriously above 1.
This causes the rapid decrease of the collision rate in QMD calculations
because of the Pauli blocking.
The behavior is completely different in the CoMD case.
The collision rate develops in about 200 fm/$c$ and reaches a
value 3 times greater than the maximum value relative to the QMD case,
which in turn gives more stopping.
After this time the disassembled system, i.e. the fragments, gradually
approach their ground state, which makes the decrease of the collision
rate.
One may notice, however, the unusually large collision rate even after
several hundreds of fm/$c$.
The reason of this large number is due to collisions
with very low relative momentum which we do not exclude in the
calculations.
If some low-momentum cut off of two-nucleon collision is put,
we could avoid those high collision rate at later times.
Nevertheless those collisions with low relative momentum have
no effect on the dynamics.

\begin{figure}
\psfig{file=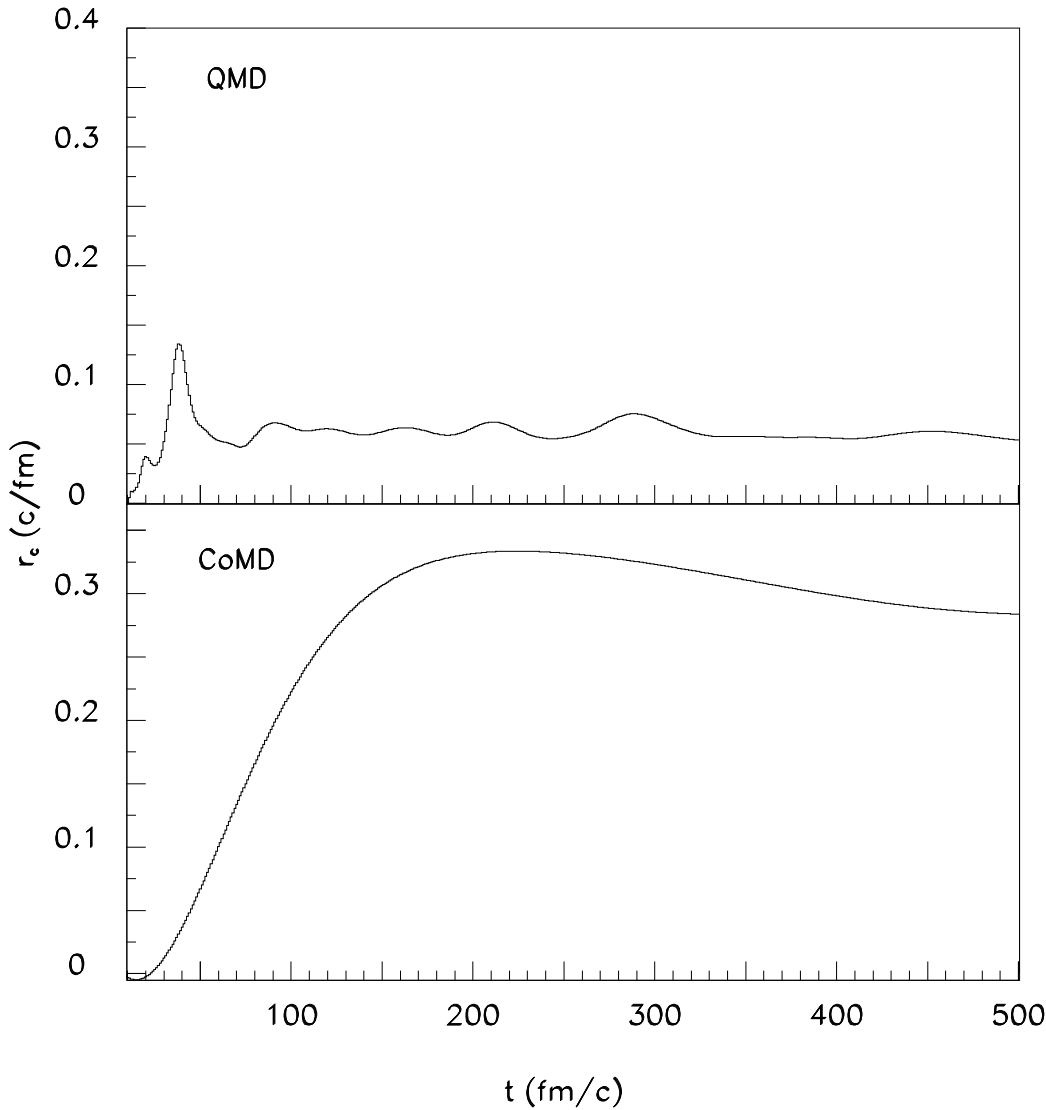,width=0.8\textwidth}
\caption{
 Collision rate $r_{c}$ as function of time for the
central collision $^{40}{\rm Ca}+^{40}{\rm Ca}$ at 35 MeV/nucleon in
the QMD and CoMD cases.
}
\end{figure}

In Figs.~4 and 5 we compare the isotope distribution
given by QMD and CoMD approaches with the experimental data on 
$^{40}{\rm Ca}+^{40}{\rm Ca}$
\cite{nat}.
About 2000 events have been generated for an impact
parameter range $b:0\sim 8$ fm.
The minimum spanning tree  method \cite{bonasera2000} in the
configuration space has been
applied to determinate the fragments at $t=300$ fm/$c$ and at
$t=3000$ fm/$c$.
We have verified that within the time 3000 fm/$c$ all the fragments are
stable.
What we can also see from the figure is that the main features of the
fragment mass distribution are already determined at 300 fm/$c$.
After that the process is dominated by the emission of
particles from the heavier fragments.

\begin{figure}
\psfig{file=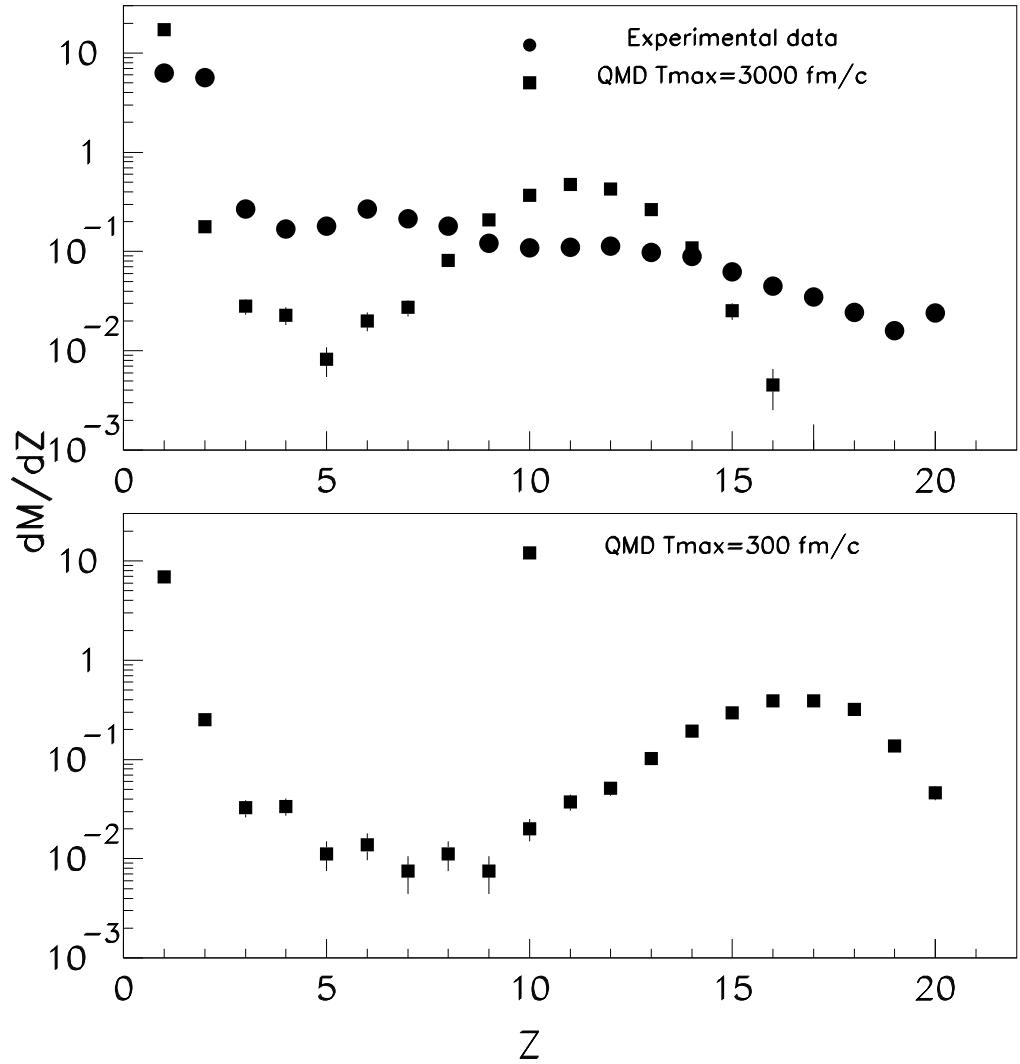,width=0.8\textwidth}
\caption{
 Comparison between the experimental isotope
distribution measured for the $^{40}{\rm Ca}+^{40}{\rm Ca}$ at 35 MeV/nucleon [10] 
system and the theoretical
prediction performed according to the QMD approach.
The calculations are shown at two different time intervals as indicated
in the figure.
}
\end{figure}

\begin{figure}
\psfig{file=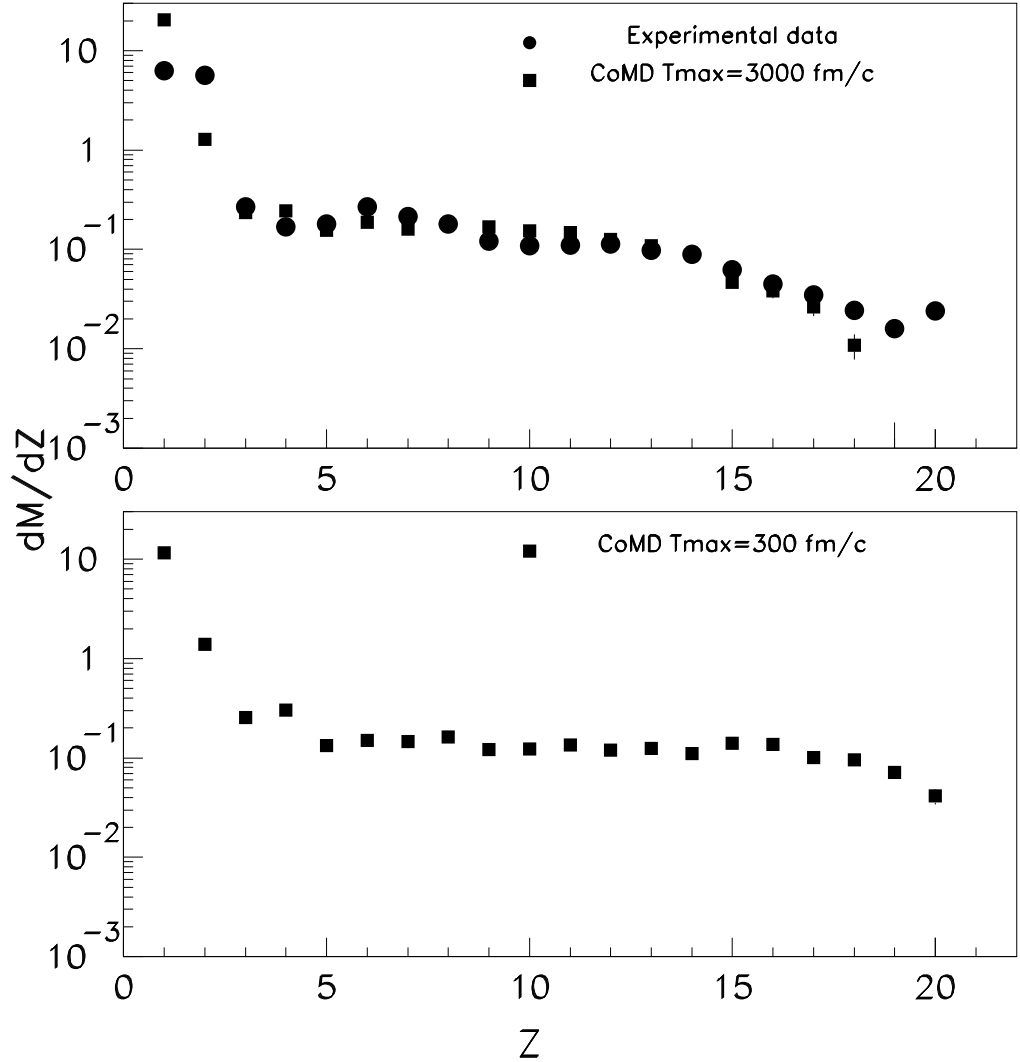,width=0.8\textwidth}
\caption{
 Same as Fig.4 but for CoMD model.
}
\end{figure}

In the upper panel of Fig.~4 we compare the experimental isotope
distribution to QMD at 3000 fm/$c$.
We observe a clear disagreement between the QMD calculation and the 
experiment.
The calculation shows essentially binary character of the reaction
while the experimental data shows a significant production of
the intermediate mass fragment (IMF: $Z\geq 3$). The binary
behavior is due to some kind of transparency
obtained in the calculations.  This in turn is due to lack of two-body
collisions because of the over crowding
of the phase space as discussed above.
At the bottom of Fig.~4 the result at 300 fm/$c$ is shown.
In this case we observe a similar behavior as above, the only difference 
being a shift of the main bump towards higher value of $Z$ and a more
pronounced ``U'' shape.
In Fig.~5 the same comparison is shown for the CoMD calculation.
The calculations are now in a satisfactory agreement
with the experimental data.
In particular  we note that the theoretical prediction of
the $Z=2$ yield is about a factor ten higher than the QMD case.
This is clearly an effect brought by the constraint (\ref{eq8})
which favors $\alpha$ particle states.
However
the predicted yields of $Z=1$ and $Z=2$ isotope
still show a marked difference with the experimental one.

We have also performed some calculations for the Au+Au system at the 
same beam energy.  Here we
are especially interested in central collision.
In fact it is quite
puzzling the behavior of such a small
system but so heavily charged.  
In Fig.~6 the charge distribution is given for impact parameters up to
3.5 fm and compared to data \cite{ref17}.
A few features are worth noticing.

1) CoMD gives too many protons (not displayed in the figure-out of scale).

2) The theoretical distribution is slightly shallower than data.  Notice
however that the data shows a jump
for $Z=20$ due to the detectors used.  If we slightly shift the data
yield up for $Z>20$ we get a better agreement to the
CoMD results.

\begin{figure}
\psfig{file=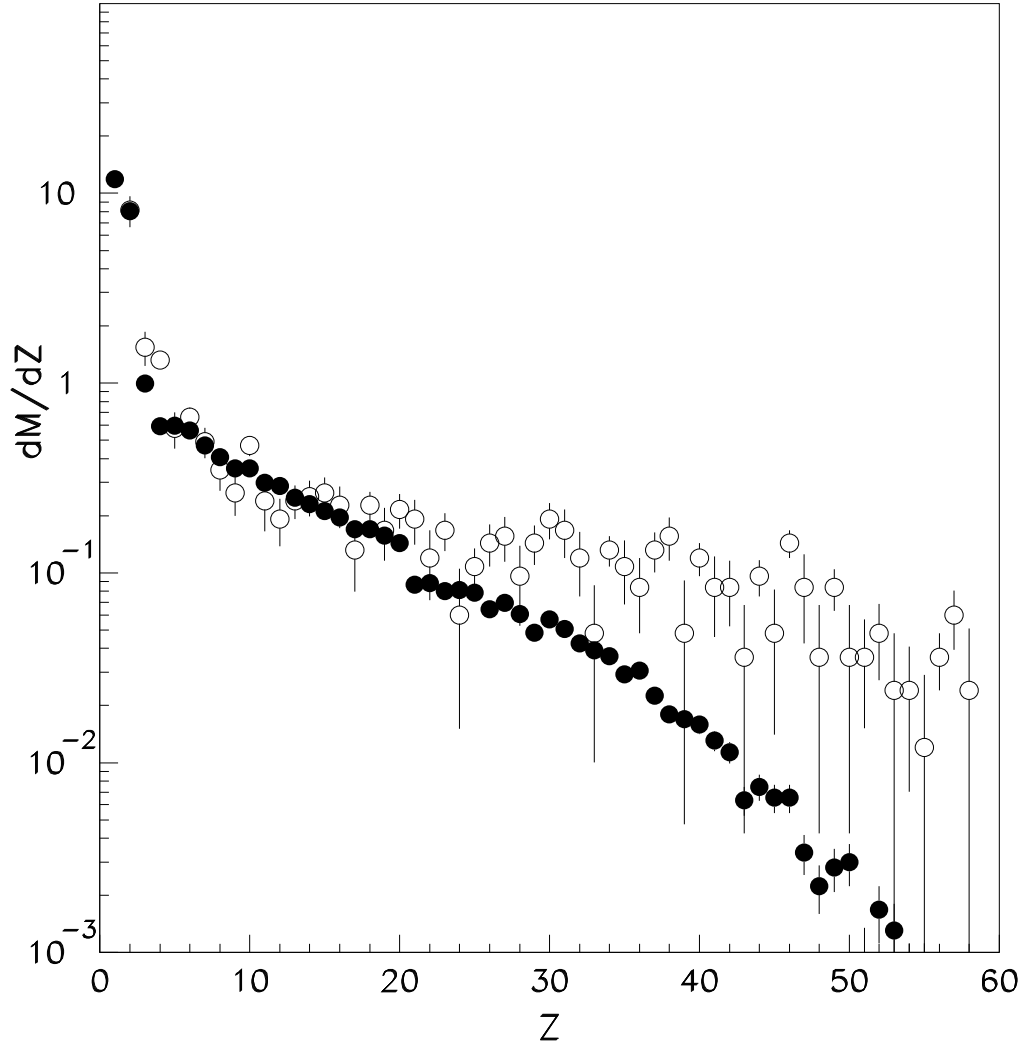,width=0.8\textwidth}
\caption{
 Comparison between the experimental isotope
distribution (full circles)[11] and the calculated one according
to the CoMD model (open circles) for central
collisions ($b\simeq 0\sim 3.5 fm$) $^{197}{\rm Au}+^{197}{\rm Au}$ at
35 MeV/nucleon.
The vertical bars indicate
the errors as due to the statistics counting.
}
\end{figure}

The system was followed up to 1500 fm/c which should be long enough to
get cold fragments.  Nevertheless we can see from 
the comparison that the model is working not too bad especially if
one recalls that other dynamical model
calculations give very steep distributions.

\section{Conclusion}

The CoMD model proposed in the present work is able 
to reproduce with the same set of 
parameters both the main characteristic of stable nuclei
in a wide region of  mass ($A=30\sim 208$) and 
the experimental isotope distribution produced in the collision
$^{40}{\rm Ca}+^{40}{\rm Ca}$ and Au+Au at $E_{\rm lab}=35$ MeV/nucleon. 
This possibility has a considerable relevance because the above mentioned 
inclusive information can not be reproduced by the QMD.
To this respect the success of the CoMD is due to the constraint 
represented by the relation (\ref{eq8}). 
This constraint, introduced to describe the Fermionic nature of 
the nuclear many-body problem, affects the dynamics of the nucleus-nucleus 
collision for two main reasons:\\
a) the nucleon-nucleon collision rate is higher with respect to the QMD case;\\
b) 
the constraint for low momentum particles  
produces obviously on the average a non local repulsion effect.\\
Both these effects play a determinant role for the disassembly of the
highly excited intermediate systems formed at the beginning of a nuclear collision 
like that investigated in this work. 
Finally we stress that for the model proposed the typical CPU time
needed to follow the time evolution of systems of mass number around 
80 for 300 fm/$c$ is quite short:
about 10 sec in a 600 MHz Unix machine.

\section{Acknowledgement}
We thank K.~Hagel and M.~Dagostino for making their data available to us.
A.B. thanks Prof.~J.~Natowitz for discussions and suggestions.
T.M. thanks INFN-LNS for warm hospitality during his stay
and M.~Colonna for fruitful discussions. Finally M.P. thanks also
several experimentalist working at the INFN-LNS for stimulating
discussions.

\end {document}